# Lithography-free Fabrication of High Quality Substrate-supported and Freestanding Graphene devices

W. Bao[1], G. Liu[1], Z. Zhao[1], H. Zhang[1], D. Yan[2], A. Deshpande[3], B.J. LeRoy[3] and C.N. Lau[1],*

[1]*Department of Physics and Astronomy,* [2]*Center for Nanoscale Science and Engineering, University of California, Riverside, CA 92521, USA*
[3] *Department of Physics, University of Arizona, Tucson, AZ, 85721, USA*

We present a lithography-free technique for fabrication of clean, high quality graphene devices. This technique is based on evaporation through hard Si shadow masks, and eliminates contaminants introduced by lithographical processes. We demonstrate that devices fabricated by this technique have significantly higher mobility values than those by standard electron beam lithography. To obtain ultra-high mobility devices, we extend this technique to fabricate suspended graphene samples with mobility as high as 120,000 cm$^2$/Vs.

* To whom correspondence should be addressed. E-mail address:lau@physics.ucr.edu

Since its experimental isolation on insulating substrates in 2004, graphene has attracted tremendous attention, as its unusual electronic, thermal and mechanical properties [1-6] promise both novel fundamental phenomena and device applications. Yet, some of the most fascinating predictions, such as Veselago lensing [7-8], fractional quantum Hall effect [9], and ballistic transistors, are yet to be experimentally demonstrated. This is partly due to the limited mobility, which ranges from ~5,000 to 50,000 $cm^2$/Vs for substrate-supported devices, and up to 250,000 $cm^2$/Vs for suspended devices [10-11]. The exact source(s) of mobility bottleneck is still under debate, but lithographical processes, which are used to fabricate almost all graphene devices to date, are known to be an important contributing factor. As graphene consists of a single atomic layer, it is particularly sensitive to surface contaminants, including resist residues left by lithographical processes, which locally modifies the electrochemical potential and provide extra scattering sites. Though annealing techniques have been demonstrated to improve device mobility [12-13], they are not well-controlled and do not always produce consistent results. Lithography-free fabrication techniques have been reported [14-15]; however, the procedures are complicated and yield devices that are restricted to simple geometries.

    Here we report a lithography-free device fabrication technique for graphene devices, via metal evaporation through silicon hard masks. This technique is simple, inexpensive, and does not require any resist processing; thus, it greatly increases device throughput, produces transparent contacts between graphene and electrodes, and yields high quality graphene devices. Additionally, hard masks, and hence devices, with complex patterns can be readily

fabricated. Using this technique, we fabricate both substrate-supported and suspended devices, whose high mobilities are characterized by electrical transport measurements.

The first and most crucial step in our fabrication procedure is the synthesis of hard silicon shadow masks, as illustrated in Fig. 1a. Here we use 500 μm double-side-polished, {100} orientation silicon wafers that are 1x1 cm$^2$ in size. Firstly, a 200-nm layer of chromium is evaporated on one side of the wafer, followed by the deposition of a thin layer of PMMA e-beam resist. This Cr layer will serve as an etching mask for later KOH and inductively coupled plasma (ICP) etching processes. Since controllable etching of thick (>100 μm) Si layer is difficult, we reduce the thickness for the final pattern etching by using photolithography and KOH etching to open a large, 400-μm deep window on the *back* of the wafer, leaving a 100-μm thick Si layer to be etched in the final step. The shadow mask structure is then patterned on the front side using e-beam lithography. After exposing and developing the resist, we use a Cr etchant (1020AC) to remove the exposed chromium layer. Finally, the shadow mask is completed by using ICP to etch through the exposed silicon layer, creating a Si wafer with patterned openings.

SEM images of two ready-to-use silicon shadow masks with different geometries are shown in Fig. 1b and c. Features as small as 500 nm can be reliably fabricated. The masks typically also contain alignment windows that assist with precison alignment during fabrication, as indicated by the arrows in Fig. 1c. These shadow masks are exceedingly robust, and can be used for more than 20 times. We note that traditional shadow masks, which consist of silicon nitride $Si_3N_4$ membranes that are partially released from Si substrates [16-19], often exhibit distorted edges[16, 18]. By comparison, our silicon shadow mask has a flat sample-contacting surface, and is sufficiently rigid for complicated structures such as Hall bars geometries.

To fabricate graphene devices, we exfoliate single layer graphene sheets on standard Si/SiO$_2$ wafers. With the help of alignment windows, we use micromanipulator XYZ translation stages to carefully align the shadow mask to identified graphene sheets, then place the entire assembly in a vacuum chamber (Fig. 1d) for metal deposition. The mask nominally rests on the substrate, though the effective mask-substrate separation, which is typically about few hundred nanometers, is determined by the thickness of the graphite residues on the substrate surface. In completed devices, we find that the metal electrodes typically extend beyond the shadow mask openings by ~ 0.3 – 0.5 µm, due to the extended size of the metal source and the finite mask-device separation.

To compare the qualities of graphene devices made by conventional e-beam lithography and shadow mask evaporation, we fabricate devices using both techniques *on the same graphene sheet*. To this end, we use shadow mask to deposit four electrodes (labeled A, B, C, D in Fig 2a), and subsequently e-beam lithography to deposit three additional electrodes (E, F and G in Fig. 2a), on a single-layer graphene sheet. The electrodes are designed to yield devices with similar aspect ratios. After *each* fabrication, the device is characterized by atomic force microscope (AFM) imaging and electrical measurements.

The right panel of Fig. 2a displays an AFM image of the graphene surface after lithography, revealing a thin layer of resist residue. The device is annealed in H$_2$/Ar atmosphere at 200 ºC for 45 minutes to remove the contaminants [12]. Using standard lock-in techniques, the two-terminal conductance of the devices are measured as a function of the back gate voltage, $V_g$, that control the density $n$ and type of the charge carriers. The device mobility $\mu$ is calculated from the slope of the $G(V_g)$ curve and the relation $\mu=\sigma/ne$, where $\sigma$ is the device conductivity, and $e$ is the electron charge. For a typical device fabricated by lithography, $\mu$ is measured to be 1500 and 3000 cm$^2$/Vs at 300K and 4K, respectively (Fig. 2c).

In contrast, for devices fabricated by shadow mask evaporation, graphene surface remains clean after evaporation, as shown by the AFM images (left panel, Fig. 2a). Atomic resolution images of the honey-comb lattice over large areas can be obtained using scanning tunneling microscopy, without any annealing treatment (Fig. 2b). From transport measurements, the device mobility is ~4000 cm$^2$/Vs at room temperature, and increases to ~7000 cm$^2$/Vs at 4K (Fig. 2d). Thus, eliminating lithography yields devices with significantly higher mobility.

This shadow mask technique can be applied to fabricate devices with a variety of geometries. As another demonstration of its power and versatility, we extend it to fabricate suspended devices via two complimentary methods. In the first technique (Fig. 3a-c), a completed device supported on a substrate is fabricated on the substrate, followed by etching by hydrofluoric acid (HF), which release the graphene sheet from the SiO$_2$ layer, and critical point drying. Here the Cr/Au electrodes double as etch masks, and HF etches anisotropically and preferentially along graphene[11], resulting in suspension of the entire graphene sheet. The HF-etched devices are annealed using current-induced Joule heating [13]. The mobility is 20,000 and 120,000 cm$^2$/Vs, respectively, at 300K and 4K. The $G(V_g)$ curves display pronounced sub-linear curvature, indicating its high mobility [10, 11] (Fig. 3d). Up to 5V can be applied to the gate voltage; above 5V, buckling of the partially suspended thin-film electrodes leads to device failure. Notably, the $G(V_g)$ curves display non-uniform variation in temperature $T$: for highly doped regimes, $G$ increases as $T$ decreases; at Dirac point, $G$ decreases with decreasing temperature. Both the mobility values and the gate-dependent $G(T)$ relation are consistent with previous measurements [11], and may also be related to formation of ripples due to graphene's negative thermal expansion coefficient [5].

For the second technique to fabricate suspended devices, a graphene sheet is directly exfoliated across pre-defined trenches on the substrate; electrodes are deposited by evaporating through shadow masks that are carefully aligned with the trenches(Fig 3e-g). The inset in Fig. 3h shows the image of a bi-layer device fabricated using this technique. Since these suspended devices do not undergo any chemical processing, they are extremely clean. Up to 30 V gate voltage can be applied to a device, since graphene is supported by the banks of the trench, not by partially suspended thin film electrodes. Device mobility of this bi-layer device is measured to be ~2000 at 300K, and 60000 cm$^2$/Vs at 4K (Fig. 3h).

In conclusion, we demonstrate a lithography-free technique for fabrication of high quality graphene devices, which may be either substrate-supported or suspended. Applications of this technique include ultra-clean devices for STM and optical measurements, or devices coupled to specialized (*e.g.* superconducting or ferromagnetic) electrodes. In particular, it provides an especially powerful approach for investigation of the mobility bottleneck of graphene devices, as it allows fabrication of ultra-clean devices that are free of both lithography contaminants and substrates.

**Acknowledgments**

We thank Feng Miao for trench wafer fabrication and Hsinyin Chiu for useful discussion.

This work is supported in part by SRC, ONR N00014-09-1-0724, ONR/DMEA H94003-09-2-0901, the U. S. Army Research Laboratory and ARO/W911NF-09-1-0333.

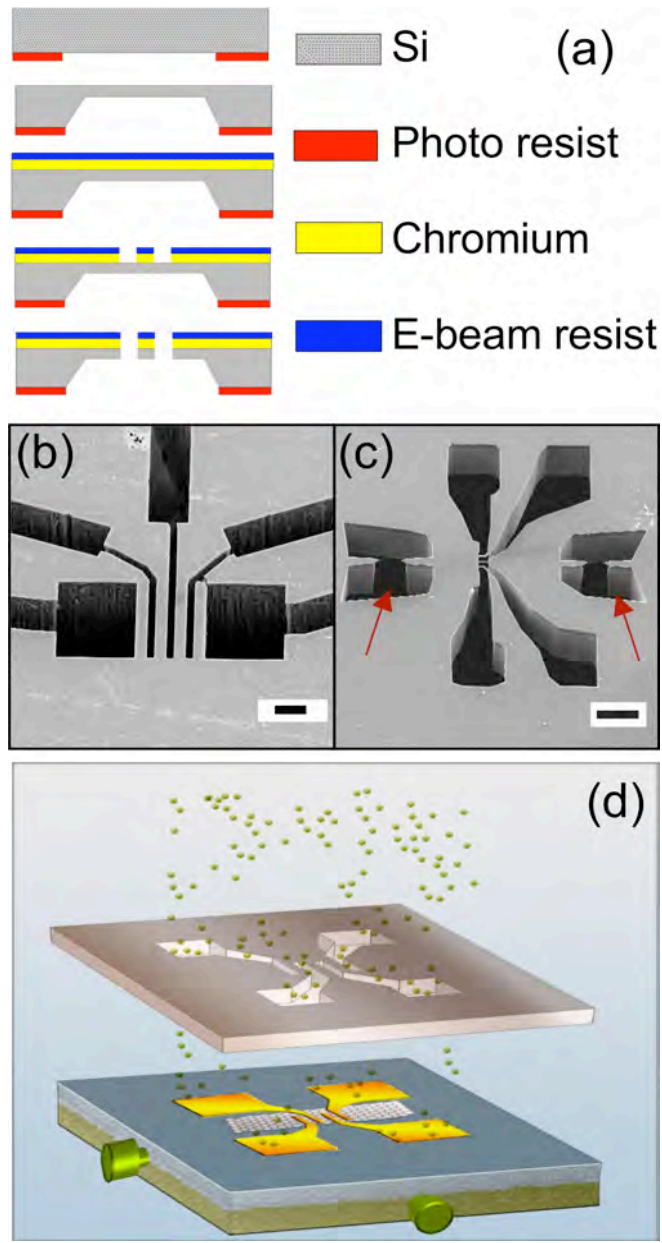

**Figure 1.** Fabrication of silicon shadow masks. (a). Schematics of the fabrication process. (b-c). SEM images of two silicon shadow masks. The red arrows in (c) indicate the alignment windows. Scale bars: 1 μm. (d). Graphene devices can be fabricated by direct deposition of metallic electrodes through these masks.

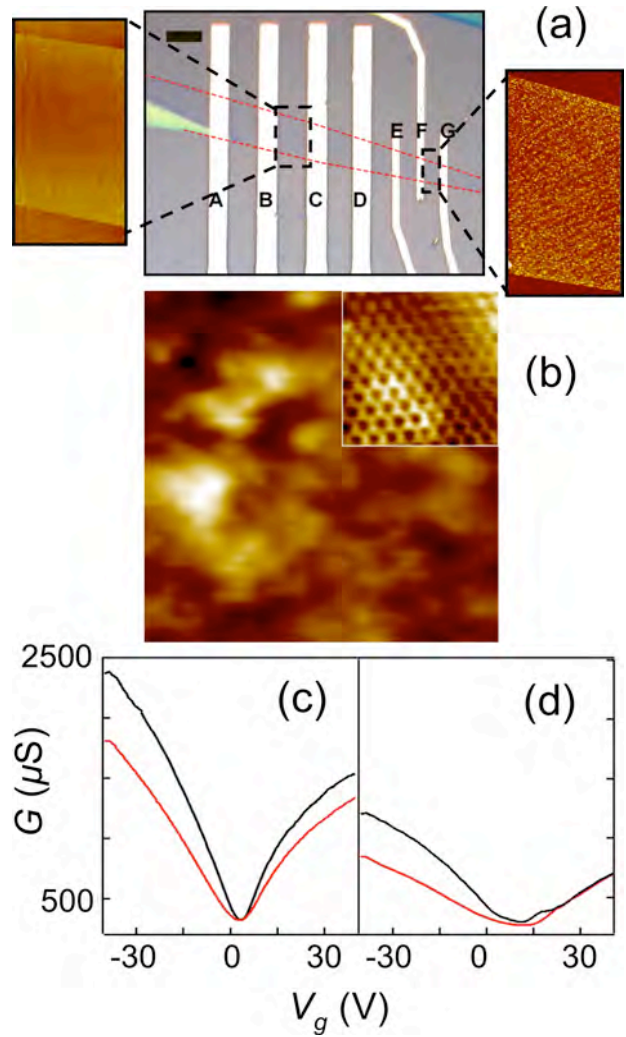

**Figure 2.** (a). Optical image of a single-layer graphene sheet device. The electrodes A, B, C and D are deposited by evaporation through a shadow mask, and E, F and G are fabricated using standard electron beam lithography. Left and right panels: AFM images of the graphene surface after shadow mask evaporation and e-beam lithography. (b). STM images of an as-fabricated device using shadow mask technique. The main panel displays an image of 85 nm x 85 nm area, and the inset shows the atomic lattice over an area of 2.5 nm x 2.5 nm. (c-d). $G$ vs, $V_g$ for the electrode pair BC and FG, respectively, at room temperature (red) and 4.2 K(black curves).

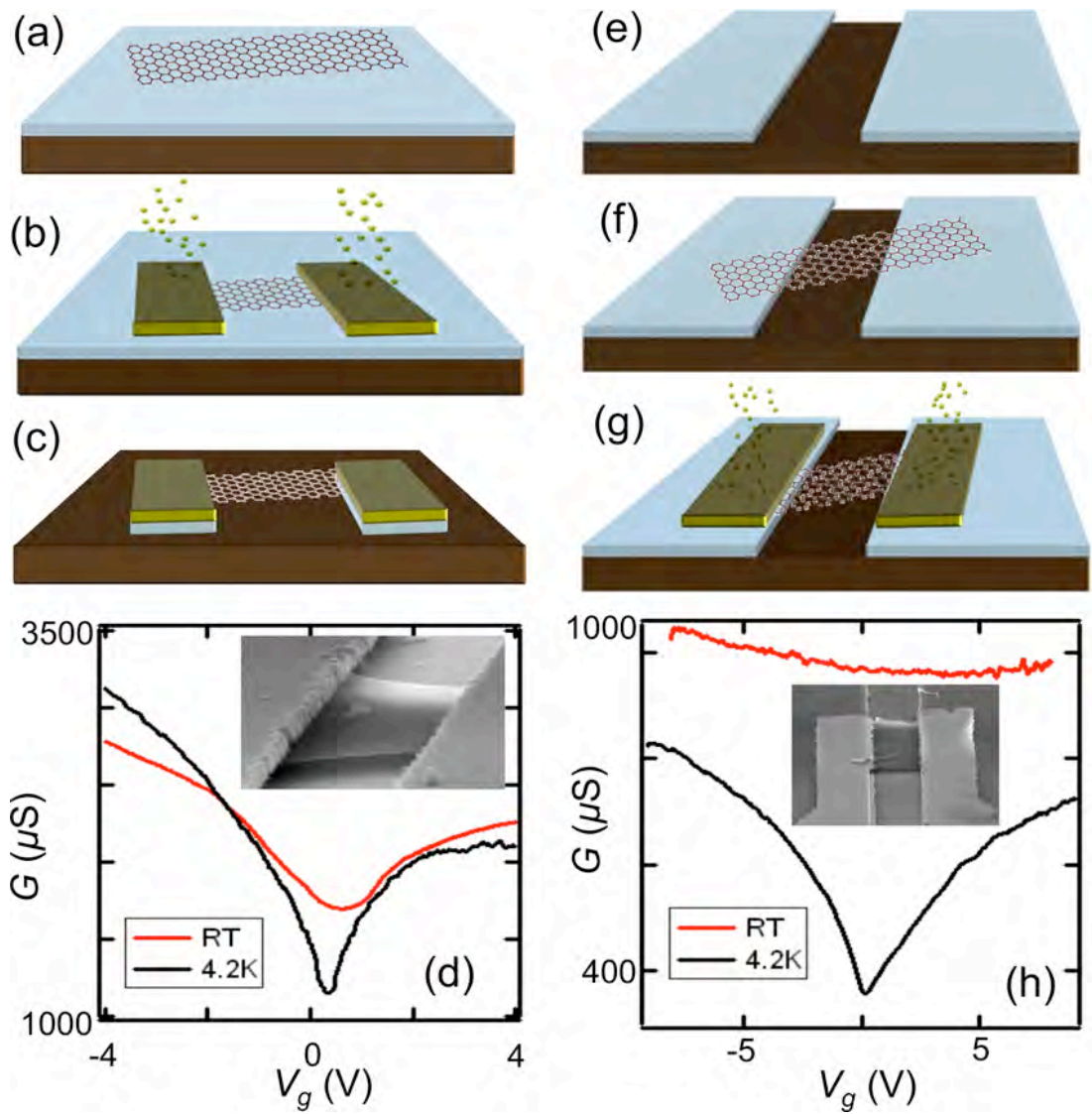

**Figure 3.** (a-c) Fabrication of suspended graphene devices via HF etching. (d). $G$ vs, $V_g$ for an HF-released single layer graphene device at room temperature (red) and 4.2K (black). Inset: SEM image of such a device. (e-g). Fabrication of suspended graphene devices over pre-defined trenches on the substrate. (h). $G$ vs, $V_g$ for a bi-layer graphene device over a trench at room temperature (red) and 4.2K (black). Inset: SEM image of such a device.